# An RFID Based Generalized Integrated System for the Identification and Traceability of Products and Subsets in Enterprises

Turcu, Cristina Elena
Prodan, Remus Catalin
Popa, Valentin

14[th] December 2005

"Stefan cel Mare" University of Suceava
Str. Universitatii nr.13, 720229 Suceava, ROMANIA
cristina@eed.usv.ro

**Abstract**

RFID tags are small electronic devices that can be used to identify objects and people. The present paper presents an RFID-based integrated system that has been developed to allow the identification and traceability of products and subsets in whatever activity field. Thus this system offers to the users the possibility to define their own information format to be stored in the transponder.
The potential beneficiaries of this system are companies that activate in fields that lend themselves admirably to RFID technologies.
Keywords: *RFID, reader, PocketPC, database..*

## 1   Introduction

Although many companies are now using sophisticated Inventory Management Systems, the identification and traceability of products and subsets is still a complex procedure which is difficult to manage.
Currently, most products and subsets tracking systems employ two-dimensional barcodes that must be close to and within the "line of sight" of the barcode reader. This is a big disadvantage that can be eliminated by manual scanning or proper positioning of the barcode and scanner. Others encountered risks using barcodes consist in non-accurate reading by the scanner caused by wet, mishandling or harsh environment. But manual scanning is labor intensive, costly, and error-prone and cannot ensure the inventory remains up-to-date, due to oversights, errors, and internal shrinkage.
RFID technology seems to be the most suitable candidate in replacing the barcodes. Once RFID technology introduced many companies began to develop dedicated applications for various domain enterprises. The main contribution of our team consists in design and implementation of a RFID based integrated system with a high degree of generalization.

## 2  RFID: concepts and benefits

RFID (Radio Frequency IDentification) is a technology used for the automatic identification of objects and people. RFID may be viewed as a means of explicitly labeling objects to facilitate their "perception" by computing devices [1]. An RFID device called RFID tag is represented by a microchip designed for wireless data transmissions. The microchip is attached to an antenna within a set upon a label. The RFID tag transmits data in response to an interrogation received from a reading device called RFID reader. Given the wireless communication between the RFID chip and the RFID reader, all data may be read from a distance. The reading range varies in accordance with the size of the reader antenna, the orientation of the RFID tag towards the antenna, the tag position with respect to the antenna core, as well as with the tag type.
Given the significant lowering of tag costs, it is expected that their usage will be considerably increased. Thus, some studies indicate that in only a few years their production will reach 1 billion or perhaps a thousand billion [2][3]. Spending on RFID implementation in the retail supply chain alone has been estimated at $91.5 million last year – an amount expected by some to exceed $1 billion by 2007 [4]. The RFID tags attached to objects may be used in supply chain management and inventory management [5].
Conceptually, bar coding and RFID are quite similar; but RFID tags have numerous advantages over barcodes. The major advantages are that RFID has the capacity to store larger amount of information, gathers the information faster than barcodes and the read/write operations can be performed through different materials such us paper, plastic or wood, with the exception of metals. RFID also allows easy and as needed uninterrupted access to data on the tag. Unlike the barcode where identification is limited by line-of-sight, RFID technology and its reliance on radio waves does not require a line-of-sight for identification nor a straight-line alignment between the tags and readers. RFID tags are also sturdier than barcodes, allowing for use in adverse conditions, and tags can be affixed or embedded on the product packaging or inside the item.
Table 1 presents a short comparison of RFID and barcodes [5]-[8]:

**Table 1. RFID versus Barcodes**

| Characteristic | RFID | Barcode |
|---|---|---|
| Reads Per Second | 40-200 | 1-2 |
| Read Range | Up to 2 m | 4-8 inches |
| Read/Write | Yes | No |
| Anti-collision capabilities (simultaneously read capabilities) | Yes | No |
| Cost | More | Less |
| Reusability | More | Less |
| Human Intervention | Less | More |
| Line of Site Required | No | Yes |
| Security | More | Less |

Speaking in enterprise terms it is evident that the usage of RFID tags in a traceability system enjoys considerable benefits [9]-[11]: high efficiency in collecting, managing, distributing and storing information on inventory, business processes, and security controls; increased productivity; products are processed at high speeds, so the time allotted to product scanning is considerably reduced; the time involved in product handling is reduced; inventory activities are simplified and data accuracy increases. Thus, various studies have proved that all inventory procedures may be performed faster than those involving barcodes [12]. Moreover, if one gets near the products holding a mobile reading system; the handheld device will immediately collect and store data; product management is improved, thanks to the re-programmable memory which also allows instant product location; customer services are considerably improved; RFID will allow receiving authorities to verify the security and authentication of shipped items.

## 3   Presentation of the system

### 3.1   General presentation

The RFID-based integrated system presents a high degree of generality and may be used for the identification and traceability of physical entities in space and time. The generalized character of the system results from the fact that it can be easily implemented in various activity fields without any modifications in the structural level of software applications. Thus, the user can define the data format to be used for writing data into tags through an advanced template editor which allows user to establish necessary fields (e.g. acquisition date, location, current value) and their type (character, string, integer, real).

## 3.2 System architecture

The architecture of the systems is presented in Figure 1.

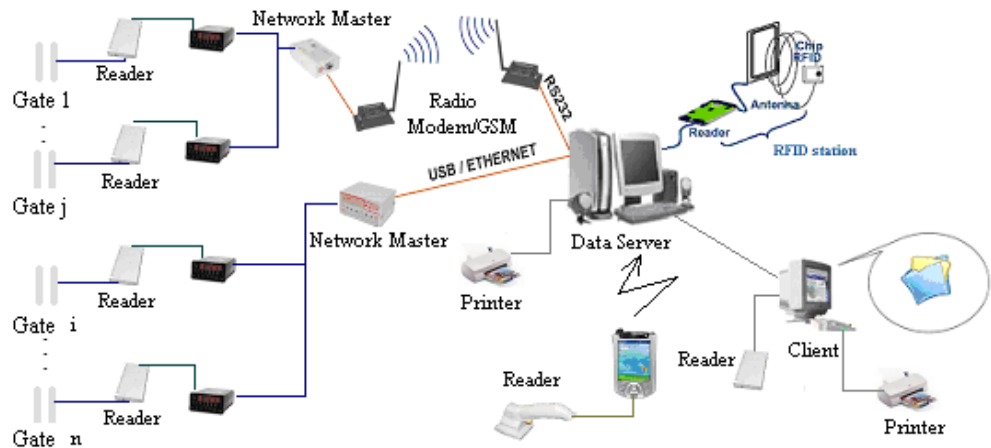

**Figure 1. System architecture**

The elements of the integrated system are:
- a PC computer running a dedicated informatics system. This informatics system must perform the following tasks:
    o to communicate with fix and mobile readers in order to retrieve data and/or transmit data to be written
    o to allow users to select the information to be introduced
    o to maintain and update a local data base that might be used for further processing. For instance, the stored data in the database can be used in complex processing activities (estimations, statistics, etc), depending on necessities.
- an RFID system connected to the PC, which allows the reading/writing of transponders;
- fix readers, formed by μPSD based stations to which readers have been connected. The stations are connected in a network, one station being able to communicate with the PC and other stations. If a great number of stations is required, a network master for maximum 30 stations will be used; this network master allows PC communication through wire or radio;
- mobile readers that comprise a connected reader to a PDA.

An RFID system is a reader and antenna combination. This allows the reading of the stored information into the RFID tag, also enabling the RFID tag update. The communication between the RFID chip and the RFID station is wireless, therefore the data can be read from a distance. Table 2 shows the technical specifications of the integrated system.

**Table 2. Technical specifications for the system**

| Characteristic | Description |
|---|---|
| Range | 1 km |
| Number of fix readers | 30 / master |
| Number of events | 255 |
| Communication | RS485 between stations |
|  | RS232 for communication with a computer (PC type) |
| Independency | the system may function independently of computer |
| Extensibility | the system may be set up within a minimal configuration, being able to be relatively easy extended, through the adding of new components |
| Tags | respects ISO 15693 standard, for 13.56 MHz frequencies |
| Reading distance | between 9 cm and 40 cm, depending on reader, antenna and transponder |
| Multiple readings | yes, the number of transponders can be simultaneously read, varying in accordance with the reader |
| Operating system | PC: Windows 2000, Windows XP<br>Pocket PC: Windows CE |

### 3.3  System facilities

Within the present integrated system, the following operations are facilitated:
- the bi-directional communication between the PC and:
  - the fix stations, using specific algorithms for wireless and radio communications;
  - mobile readers, allowing a total or a partial transfer of registrations within database tables from PC to PDA and in inverse order;
  - reader, enabling the reading of stored information into transponders and the update of databases from the PC, as well as the writing and updating of transponders with information according to the settings performed by the user;
- enabling the communication between the PDA and:
  - the PC, for the update of the database from the PC and from the PDA, respectively;
  - reader, for achieving transponders reading and database updating from the PDA, respectively the writing and updating of transponders in accordance with the registrations in the database.

- enabling the communication between the station and:
    - the PC, for sending the commands to the station, respectively data base update from the PC with station read information or with the events generated by stations;
    - reader, for carrying out the reading and respectively writing of transponders existing within the reader range, depending on the commands conveyed to the stations;
- enabling the system security, through system access control, as well as data encrypting from the database. These operations are performed both on the PC and the PDA;
- management of system registered users (users visualization, adding or deletion of certain users, profile modification, etc.);
- enabling the management of the database, which stores information related to transponders, station control, events derived from system stations, users and events;
- the configuration of the integrated system in accordance with necessities, including the configuration of:
    - the applications running on the PC, with the setting of the folder holding the application database, the station communication rate, displaying color (Figure 2);
    - the working stations, allowing, for instance, the setting of the communication rate, station address, password, etc.;
    - PDA application, with the setting of application data basis folder.
- warning/alarming overrun expected/special situations, in accordance with the settings;
- creating the models to be used for transponder information storage (Figure 3);
- administration of databases with information on transponders, stations, etc.;
- enabling the communication with system customers, and allowing the connection with higher enterprise levels;
- the storage of all events and their administration (visualization, searching, filtering etc. – Figure 4, 5);

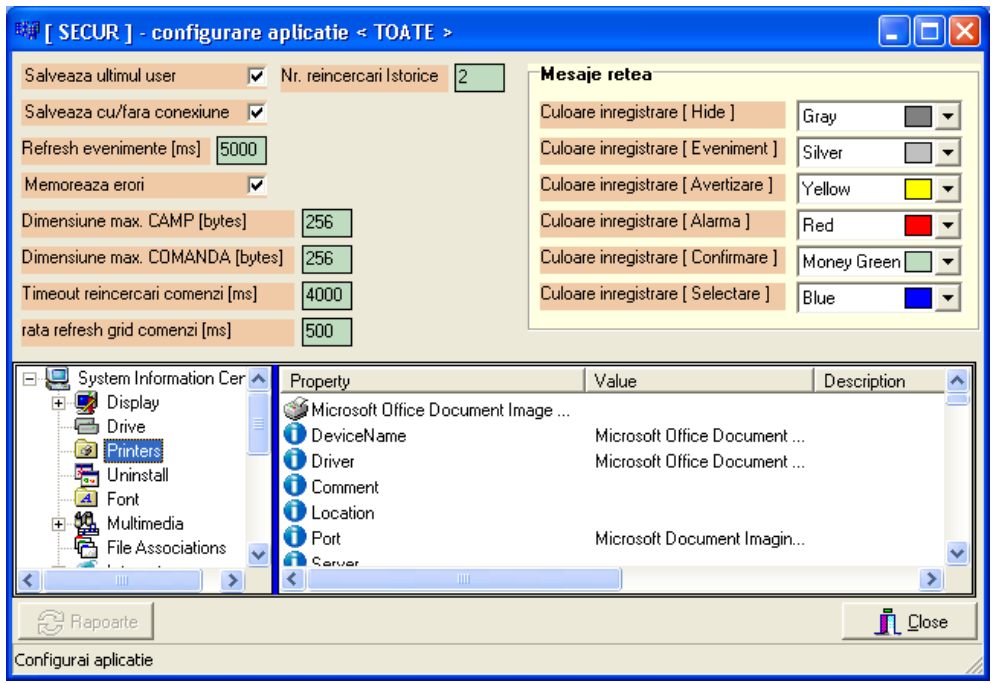

**Figure 2. Example of a configuring window**

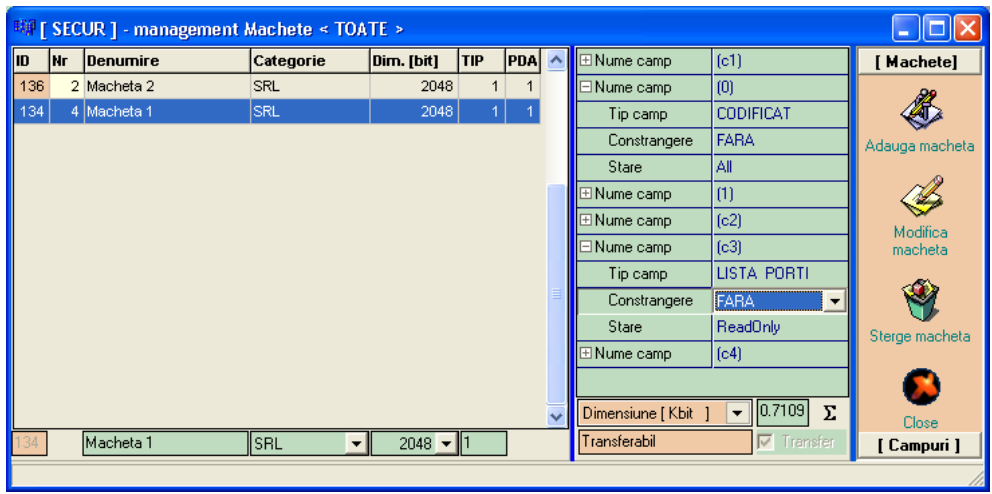

**Figure 3. Example of a window used in models achievement**

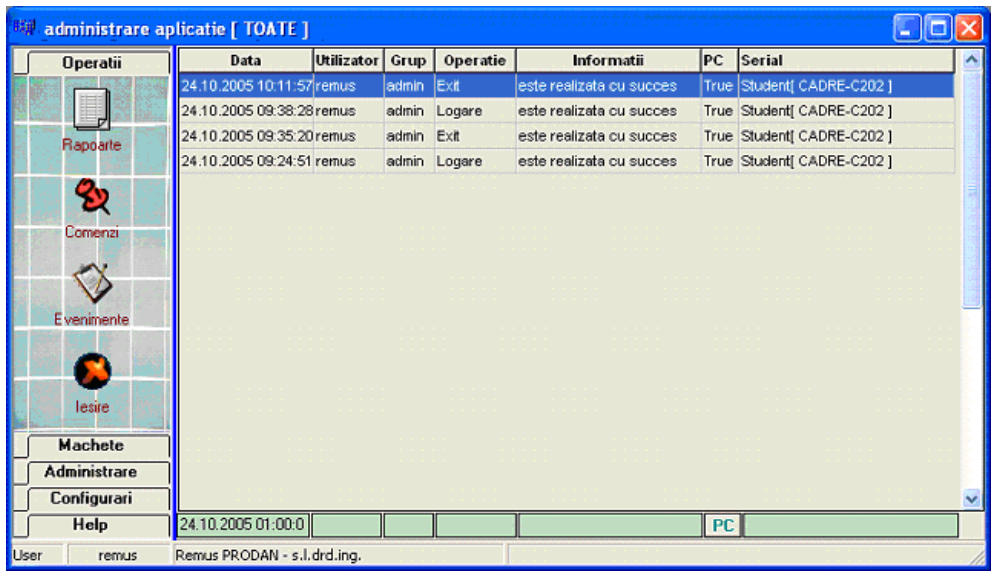

**Figure 4. Window with a PC events journal**

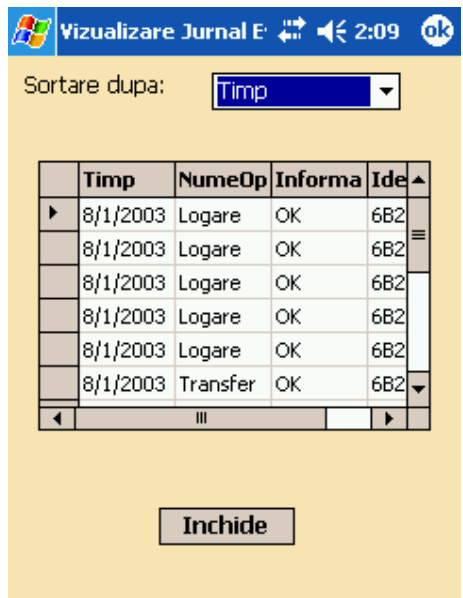

**Figure 5. Displaying of PDA events journal**

- the visualization and processing of interesting information;

- the defining of some report patterns in accordance with the user's options and the simple generation of reports, with the possibility of their visualization and printing, (Figure 6)

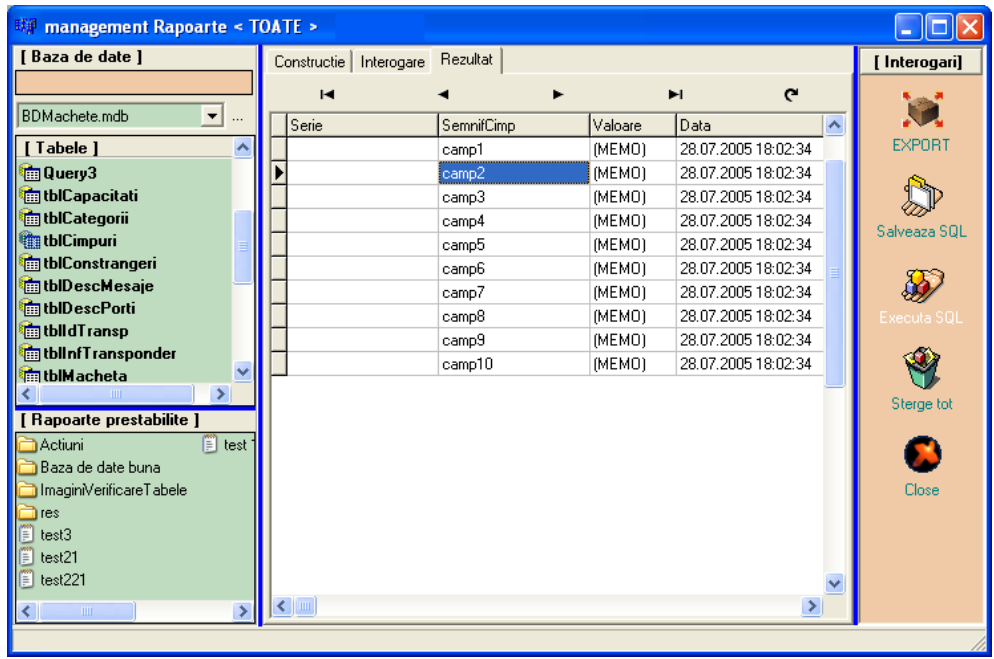

**Figure 6. Reports management**

As for the database format, the Microsoft Access format (.mdb) was chosen for the PC and the Pocket Access format (.cdb) for the PDA. The PC<->PDA communication is commonly carried out through the agency of Microsoft ActiveSync application, which has the function of synchronizing one or more files between the PDA and the PC. Irrespective of the database format of large files, the transfer carried out by Microsoft ActiveSync takes a long time because the entire database is being transferred.

Within the present system, only the tables carrying modifications are thought to be updated. The design and implementation of a functions library was necessary to solve conversion problems of database files from the PC to the PDA and vice versa. This library ensures:
- the establishment of the PC-PDA connection; the server application will ensure the connection to the PDA;
- the PC-PDA connection interruption;
- the PC-PDA connection state ascertainment;

the processing operations on files/folders: copying of a certain file from the PDA to the PC, copying of a certain file from the PC to the PDA, deletion of a certain file from the PDA, creating a new folder on the PDA, deletion a folder from the PDA, creating a new folder on the PDA, deletion of a folder from the PDA, retrieving information about a file/folder, copying a Microsoft Access database (.MDB) from the PC to the PDA and the conversion to Pocket Access format (.CDB) and vice versa, copying of certain tables from a Microsoft Access database (.MDB) from the PC to the PDA and adding them to a Pocket Access database (.CDB) and vice versa.

## 4   Potential beneficiaries

The potential beneficiaries of this system are companies that activate in fields that lend themselves admirably to RFID technologies:
- supermarkets, stores, bookshops, pharmacies etc – where the entities are the products on sale; these products will no longer carry barcodes.
- airports – where the monitored entities are the passengers' luggage;
- libraries – where transponders will facilitate the traceability of both library books and registered readers;
- manufacturing companies – where the entities range from semi-finished goods and subsets to end products; a traceability systems will allow the perfect management of all sets and subsets.
- various enterprises wishing to keep their stock in sight and to make fast inventories.

## Conclusions

RFID is sometimes named a revolution for the identification and traceability of products and subsets in an enterprise. The benefits that RFID can bring are numerous, but so are the challenges that come with it.
The paper presented an integrated system based on RFID technologies characterized by a high degree of generality, which makes it perfect for the identification and traceability of various physical entities in space and time.
The developed system may be easily adjusted to users' requests, ensuring high performance and flexibility. The user graphics interface is simple to use and allows varied configurations depending on user preferences and necessities. The usage of the present system results in a considerable reduction of human errors. It also promotes quality and ensures high-speed data processing. Requiring no software modifications, the system is ideal for extremely varied activity fields. The potential beneficiaries of this system are companies that activate in fields that lend themselves admirably to RFID technologies.